# Method of Controlling Corona Effects and Breakdown Voltage of Small Air Gaps Stressed by Impulse Voltages

Athanasios Maglaras[#1], Trifon Kousiouris[*2], Frangiskos Topalis[*3], Dimitrios Katsaros[$4],

Leandros A. Maglaras[$4], Konstantina Giannakopoulou [#5]

[#]*Electrical Engineering Department,*
*T.E.I. of Larissa, 41110 Larissa, Greece*
maglaras@teilar.gr
giannakopoulou@teilar.gr

[*]*Electrical and Computer Engineering Department, N.T.U.A.*
*9, Iroon Polytechniou str.*
*157 80 Athens, Greece*
tkous@softlab.ntua.gr
topalis@ieee.org

[$]*Computers, Telecommunications and Networks Engineering Department, University of Thessaly*
*37 Glavani – 28th October Str*
*Deligiorgi Building, 382 21 Volos, Greece*
dkatsar@inf.uth.gr
leadrosmag@gmail.com

*Abstract—* **This paper investigates the influence of a resistor on the dielectric behavior of an air gap. The resistor is connected in series with the air gap and the latter is stressed by impulse voltage. Air gap arrangements of different geometry with either the rod or the plate grounded are stressed with impulse voltages of both positive and negative polarity. The resistor is connected in series with the air gap in the return circuit connecting the gap with the impulse generator. The method followed involves the investigation of the graphs of the charging time concerning the air gaps capacitances, in connection to the value of the resistor, the geometry of the gap, the effect of grounding and the polarity effect. It is determined that the charging time of the air gap increases, as the value of the resistor increases. It is also determined that the peak voltage value of the fully charged air gap decreases as the value of the resistor increases. The results of the mathematical and simulation analysis are compared with the results of the oscillograms taken from experimental work. In addition and consequently to the above results it is concluded from the experimental work that the in series connection of the resistor in the circuit has significant influence on corona pulses (partial discharges) occurring in the gap and on the breakdown voltage of the gap. A new method of controlling the corona effects and consequently the breakdown voltage of small air gaps stressed by impulse voltage of short duration in connection to the ground effect and the polarity effect has arisen. Furthermore through mathematical analysis of the charging graphs obtained from simulation and experimental oscillograms there was a calculation of the values of the capacitance of the air gaps in relation to their geometry and the results were compared to the values calculated with mathematical analysis.**

## INTRODUCTION

Air as an insulator is the most used in various arrangements and probably the best conventional solution for the most of the high voltage applications. The air gap thus is considered as one of the most important parameters for the design and dimensioning of insulating arrangements, in almost every electrotechnical application.

In designing nearly every electrical arrangement, air gaps are essential components that arise necessarily in constructions (switches, gaps between power lines, or power lines and earth, gaps between electrical and electronic components in most devices, etc.), and they are stressed by dc ac or impulse voltages.

The basic effects which are referred as the dielectric behavior of an air gap are the corona effects and the breakdown voltage, [1-3]. The basic magnitudes which describe the dielectric behavior of an air gap are the corona onset voltage, the corona current or pulses, the breakdown voltage, and dielectric strength, [1-6].

The most known effects which influence the values of the above mentioned magnitudes are, the polarity effect, [1-3], [7-10] and the barrier effect. Other lately investigated phenomena which have great influence on the dielectric behavior of the air gaps are the ground effect, that is the influence of the different electrode of the gap chosen to be grounded on the field distribution and hence the dielectric behavior of a gap [11-13], and the corona current effect, that is the influence of the corona current on the dc breakdown voltage of an air gap.

Impulse voltage is the form of voltage that describes the voltage obtained on a construction struck by lightning or on a power line when a switch is opened or closed under power. It is a one polarity pulse of short total duration, with a very short rise time and a much longer fall time.

Impulse voltages influence the insulation materials and the air gap differently in comparison to dc and ac voltages. Their duration is very small and in some cases there is not enough time for the effects of corona, partial discharges and breakdown to occur, because of the necessary time lag.

The corona effects which occur in air gaps stressed by dc or low frequency ac voltages before breakdown postpone the breakdown mechanism, while with impulse voltages they lead to breakdown.

Generators producing impulse voltages of different time duration have been used for many years mainly for testing the behavior of the insulating arrangements. In Fig. 1 a schematic diagram of αn impulse generator of one stage is shown.

Air gap arrangements have been investigated under impulse voltages for many decades. The mostly studied air gaps are the sphere-sphere, rod-rod and especially the rod-plate or point - plate air gaps mainly when the plate is grounded, because these gaps feature the least dielectric strength and the most intense corona effects, especially when the plate is grounded. Air gaps with the rod grounded stressed by impulse voltages have not been investigated yet. In such arrangements the electric field in the gap is less inhomogeneous and the voltage needed for the corona pulses to occur is higher and consequently the breakdown voltage is also higher. Perhaps the first publication that provides a detailed analytical and experimental approach for the arrangements with the rod grounded, with significant results is [11]. The effect is called the ground effect and as it is presented in this paper it influences greatly the air gaps stressed by impulse voltages, as well.

It is well known that in a stressed air gap arrangement, the polarity of the rod has a great influence on the corona effects [1-3]. When the rod is negative in comparison to the plate the corona onset voltage is lower and the corona effects more intense, [12]. This is the polarity effect and it is known to influence the dielectric behavior of air gaps stressed by impulse voltages, [1].

In almost every air gap arrangement investigated, one electrode is grounded (mainly the plate) with a ground resistance less than an Ohm. Arrangements with a relatively big resistance connected in series with the air gap have not been investigated yet. The present paper investigates the influence of a resistor connected in series with the air gap stressed by impulse voltages on the dielectric behavior of a gap, in correlation with the ground effect and the polarity effect. It is a new way of air gap circuit connection that leads to new results concerning the corona effects (values of the corona onset voltage and corona pulses) and the values of the breakdown impulse voltage. The results depend also on the ground effect (electrode chosen to be grounded) in combination to the polarity effect.

Furthermore the used method leads to a calculation of the values of the capacitance of the air gaps arrangements.

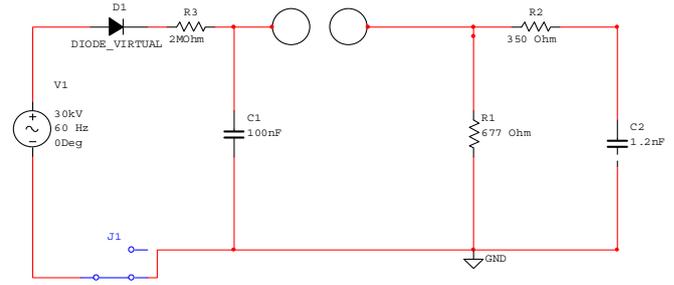

Fig. 1 The schematic diagram of the impulse generator.

In air gaps stressed by impulse voltages the corona effects occurred before breakdown with the form of current pulses produce electric charges in the gap and create enhanced conditions for the emergence of avalanches leading to the breakdown of the gap. As corona pulses become more intense the values of the breakdown voltage decrease [1]. The corona effects and the breakdown of rod-plate air gaps have been experimentally investigated by many researchers for the most commonly used arrangements where the plate electrode is grounded and voltage is applied to the electrode of the rod [4-5], [7-9].

Several methods have been proposed for controlling the breakdown voltage, when stressed by dc or ac voltage, [2-3-11]. In the present paper a new method is investigated based on the results given from the influence of a resistor connected in series with an air gap stressed by lightning impulse voltage in combination with the effect of grounding and the polarity effect.

THE PROCEDURE FOLLOWED

Experimental tests on rod-plate air gaps stressed by impulse voltage have been fulfilled with the experimental set up as schematically illustrated in Fig. 1. This arrangement has been built up to measure the applied to the air gap impulse voltage, and the value of the voltage $V_{R4}$ across the resistor $R_4$, as well as the corona pulses and the breakdown voltage occurring in the gap. The corona current through the gap can be easily calculated by the following equation:

$$I_{gap} = V_{R4}/R_4 \qquad (1)$$

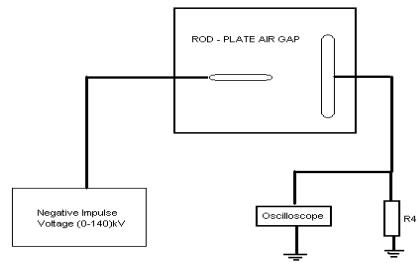

Fig. 2 The experimental arrangement schematically.

There are four basic characteristics that define the parameters of an impulse voltage. These are front, or rise time

($t_1$), tail or fall time or a time to half-value ($t_2$), peak value of the voltage ($V_{imp-peak}$) and the polarity. At the same time the form of the impulse voltage is very important. The test systems generate lightning impulse voltage (LI, 1.2/50 μs), and switching impulse voltage (SI, 250/2500 μs) in accordance with IEC 60060-1.

Impulse voltages are produced by impulse generators. The schematic diagram of the impulse generator used in the present paper is shown in Fig. 3. In used model $C_1$=100 nF, $C_2$=1.2 nF, $R_1$=677 Ohms, and $R_2$=350 Ohms, while the value of air gap capacitance ($C_{gap}$) is a few pF and the value of the ground resistance ($R_g$) is less than an Ohm.

The maximum value of the output voltage of an impulse generator applied on the air gap well grounded is usually a little smaller than the voltage of the $C_1$ capacitor, the difference depending on the circuit components and the load of the generator.

The charging time of the air gap connected to a power supply is given by equation T=k.R.$C_{gap}$, where R is the resistor through which the gap is charged, $C_{gap}$ is the capacitance of the gap and k is a coefficient, the value of which depends on the circuit data and desired accuracy. In an air gap arrangement the capacitance is very small and when it is connected to the generator with a negligible ground resistor, and $R_4$ is very small (Fig. 3), the charging time of the capacitance of the gap arrangement is very little compared to the rise time of the impulse voltage (1.2 μs) and the maximum voltage value of the fully charged gap ($V_{gap}$) is the same with the maximum value of the applied impulse voltage. On the contrary when the value of the resistor $R_4$ is high the charging time may become much bigger than the rise time of the applied impulse voltage and so the final voltage of the fully charged gap ($V_{gap}$) may be significantly lower than the maximum value of the applied impulse voltage (Fig. 4), because the end of the charging time is located in the region that coincides with the fall time of the impulse voltage. And as expected the lower value of the gap voltage decreases the likelihood of corona effects or breakdown to occur subsequently after.

Corresponding results arise when a second air gap same with the investigated one is connected in series to it, or perhaps a proper capacitor of very small capacitance, comparable to the one of the air gap.

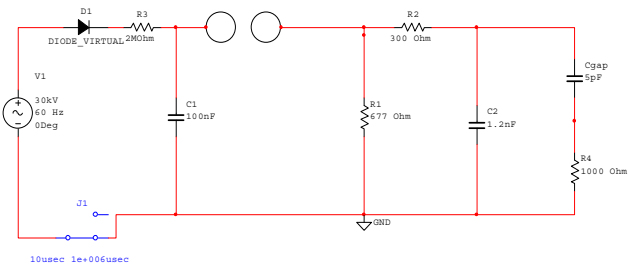

Fig. 3 The schematic diagram of the experimental arrangement. $C_{gap}$ is the capacitance of the air gap arrangement

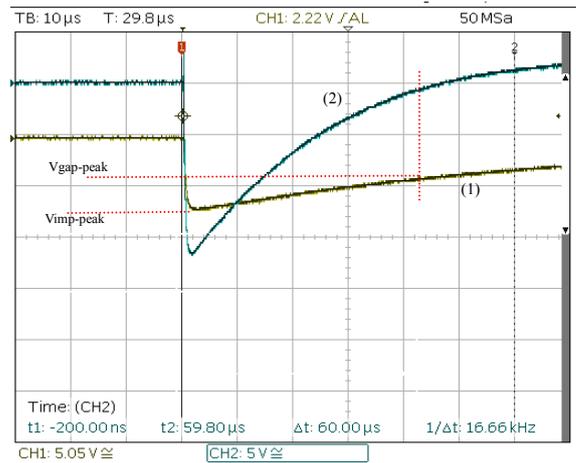

20 kV/div, 10 μs/div, (1) and (2) are the graphs of the impulse applied voltage ($V_{C2}$) and the voltage across $R_4$ respectively.
Fig. 4 Oscillogram showing the decrease of the maximum voltage of the fully charged rod-plate air gap when the R4=100 kOhms. Charging time of the gap arrangement is 47 μs, the maximum value of the impulse applied voltage is 30 kV (1), while maximum value of the fully charged gap is 16 kV (2).

The resistor $R_4$ is connected in series with the air gap in the return circuit before ground (Fig. 3). The method followed was to monitor and investigate the graphs of the applied impulse voltage and the voltage form across $R_4$ ($V_{R4}$) and calculate the charging time concerning the air gaps capacitances, and the maximum voltage of the fully charged gap (Fig. 4). The value of the impulse voltage used was relatively small so that corona pulses or breakdown should not occur.

Models of the experimental arrangements have been drawn and analyzed by simulation and mathematically, and the results were compared with the results of the oscillograms and TRC data taken from the oscilloscope of the experimental set up.

Furthermore through mathematical analysis of the charging graphs obtained from simulation and experimental oscillograms there was a calculation of the values of the capacitance of the air gaps in relation to their geometry.

In addition and consequently to the above results the influence of the in series connection of the resistor on the corona effects and breakdown voltage of the gap is experimentally investigated.

## EXPERIMENTAL RESULTS

The experimental arrangement used (Fig. 1) is consisted of an impulse generator capable of producing lightning impulse voltage with duration 1,2/50 μs and value up to 140 kV, a special test cell construction with air gaps of different geometry and resistors ($R_4$) from 120 Ohms to 100 kOhms. Digital storage oscilloscope (Model HAMEG, HMO1022, 100 MHz) has been used to detect and monitor the voltage across capacitor $C_2$ (output voltage of the impulse generator) as well as the voltage across resistor $R_4$ ($V_{R4}$). The air gap charging current through the cap can be easily calculated from equation (1).

The arrangements, which have been experimentally studied, are typical rod-plate air gap arrangements of different electrode geometry and gap length connected to a lightning impulse generator. The rod electrode is a hemispherical capped long cylinder with a relatively small diameter (4 or 10 mm), and the plate electrode is a disk of 100 mm in diameter, both made of brass. Impulse voltage of negative or positive polarity is applied to one electrode while the other is at earth potential (grounded).

The influence of the surrounding is minimized, by keeping relatively big distances between the models and the boundary shielding, as well as between the experimental arrangements and the grounded elements of the laboratory. Anyway the parasitic capacitances that exist are added to the capacitance of the air gap leading to a differentiation of the test results. These capacitances can be separately measured and taken in to account.

## A. The Influence of Ground Effect and Polarity Effect

Using a very small resistor $R_4$ (a few Ohms) the graphs of the corona pulse and breakdown voltage for different arrangements of rod-plate air gaps with the plate or the rod grounded have been monitored. The V50% breakdown lightning impulse voltages of the air gaps have also been measured.

The measured values of the corona pulses and breakdown voltage of rod-plate air gaps were used in the graphs of Figs. 5 and 6. It is obvious that the ground effect (plate or rod grounded) and the polarity effect (positive or negative polarity of the voltage) influence greatly the dielectric behavior of air gaps.

When the rod-plate air gaps are stressed by lightning impulse voltages the results are different from the results recorded with dc voltage [11]. With dc voltages the occurred corona current (average value of the corona pulses) increases the value of the breakdown voltage [12]. When impulse voltage is applied to the gap the corona effects appear in a form of one or rarely two partial discharge pulses of very short duration that occur before breakdown, and they greatly influence the values of the V50% of the breakdown voltage of the gap. When the corona pulses are intense the values of the breakdown voltage decrease [1].

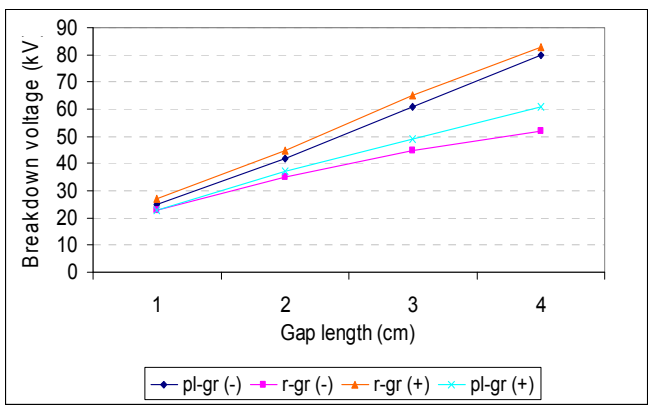

Fig. 5 The values of the breakdown voltage (V50%) of rod-plate 10-100 mm air gaps, 3 cm in length, stressed by impulse voltage 1,2/50 μs in combination with the ground effect and the polarity effect.

It can be concluded that higher values of breakdown voltage appear in the arrangements with the rod grounded stressed by positive impulse voltage, or the plate grounded stressed by negative voltage (arrangements with negative rod), while lower values appear in the rod grounded gaps stressed by negative voltage, or the plate grounded stressed by positive voltage (arrangements with positive rod).

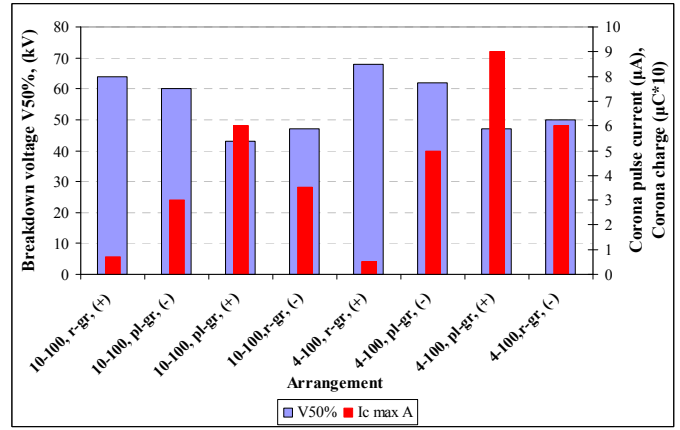

Fig. 6 The values of the breakdown voltage of rod-plate 10-100 mm air gaps, 3 cm in length, stressed by impulse voltage 1,2/50 μs of both polarities, in comparison to the peak values of the corona pulses occurred before breakdown.

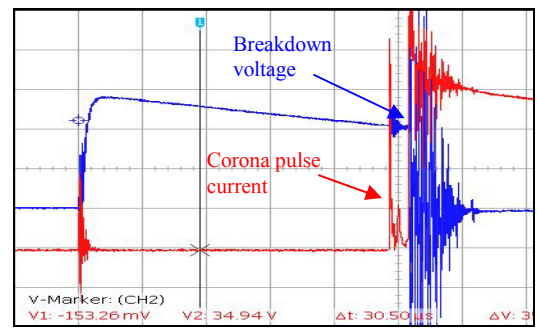

(a) Plate grounded air gap, large corona pulse, and lower breakdown voltage. 20 kV/div, 0.8 A/div, 5μs/div

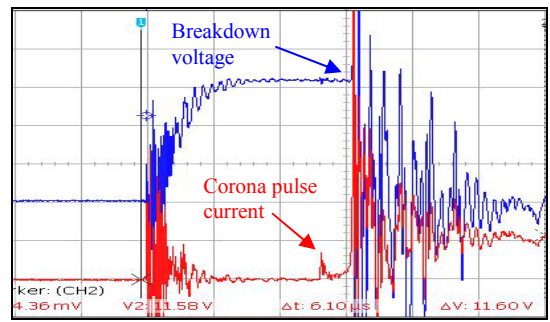

(b) Rod grounded air gap, small corona pulse and higher breakdown voltage. 20 kV/div, 0.8 A/div, 1μs/div

Fig. 7 Oscillograms showing the breakdown positive impulse voltage and the corona pulse occurred before breakdown for the 10-100 mm rod-plate air gaps, with a length of 3 cm. $R_4$=120 Ohms.

From Fig. 7 it can be concluded that the maximum values of breakdown impulse voltage occur in the plate grounded arrangement stressed by impulse (+), in which the minimum values of corona pulses are recorded, while the minimum values of the breakdown voltage occur in the plate grounded arrangements stressed by impulse (+) in which the maximum values of the corona pulses are recorded. The differences depend on the intensity of the corona effects in combination with the ground effect and the gap's geometry and are bigger in the arrangements with the rod grounded, with smaller rod's diameter.

In the rod-grounded arrangements the field is less inhomogeneous and hence the values of the corona onset voltage are higher, the corona pulses smaller and the breakdown voltage higher.

In air gaps with smaller diameter of the rod (4 mm) the critical volume is smaller than in the ones with bigger diameter (10 mm), and hence the values of the corona onset voltage and the breakdown voltage are higher, but since the field is more inhomogeneous the corona effects are more intense, and the corona pulses higher.

*B. The Influence of a Resistor in Series with the Gap*

Using resistors $R_4$ of higher values (kOhms) and relatively low values of lighting impulse voltages, so that no corona or breakdown occur, the graphs of the voltage across the resistor $R_4$ were monitored, consequently and in comparison to the graphs of the output voltage of the impulse generator ($V_{C2}$). These graphs present the gaps capacitance charging curve, and hence it is easy for the charging time, the charging maximum current, and the final maximum voltage value of the fully charged gap to be measured or calculated (Fig. 4).

It is resulted that the charging time ($t_{ch}$) of the air gap (measured as voltage across the resistor $R_4$) increases as the value of the resistor $R_4$ increases, and is different for different geometry of the gaps (Fig. 8). It is also resulted from Fig. 9 that the peak voltage value ($V_{gap-peak}$) of the fully charged air gap decreases as the value of the resistor increases, the difference depending on the form of the applied impulse voltage and the geometry of the arrangement.

The capacitance of the gap can be mathematically calculated from the curve of the voltage $V_{R4}$, or better from the TRC data of the oscilloscope. The current through the gap is calculated by equation (1), the total electric charge of the fully charged gap by equation:

$$Q_{gap} = \int_0^{t_1} I(t) \cdot dt \quad (2)$$

and the Capacitance of the gap arrangement from equation:

$$C_{gap} = Q_{gap} / V_{gap-peak} \quad (3)$$

Since the maximum final charging voltage of the gap is significantly smaller than the applied maximum value of the impulse voltage, it can be easily concluded that the value of the impulse voltage needed for the corona effects and the breakdown voltage to occur increases.

It has been proved through experimental procedure that for the same value of impulse voltage applied the probability for the air gap to breakdown minimizes for bigger values of the resistor connected in series with it. It is also resulted that the forms of the corona pulses are different for different values of the resistor $R_4$.

Table 1 features the test results for a rod - plate air gap with 10 mm and 100 mm diameters respectively and a length of 4 cm, stressed by negative lightning impulse voltage for different values of resistor $R_4$.

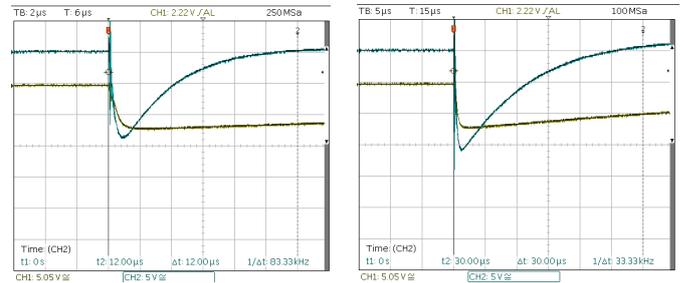

(a) $R_4$=12 kOhms Charg. time 10.5 μs   (b) $R_4$= 33 kOhms charg. time 25 μs
  20 kV/div, 2 μs/div                       20 kV/div, 5 μs/div

Fig. 8 Oscillograms showing the impulse voltage (1) and the voltage across $R_4$ ($V_{R4}$) for the 10-100 mm rod-plate air gaps, with the plate grounded and a length of 4 cm, stressed by 30 kV lighting impulse voltages.

TABLE I
BREAKDOWN PROBABILITY OF ROD-PLATE AIR GAPS FOR DIFFERENT VALUES OF THE RESISTOR $R_4$, IN CONNECTION TO THE GROUND EFFECT.

| Arrangement | Voltage on $C_2$ (kV) | $R_4$ (kOhms) | Breakdown Probability (%) |
|---|---|---|---|
| Plate grounded | 84.5 | 0.12 | 100 |
|  |  | 1 | 90 |
|  |  | 10 | 70 |
|  |  | 100 | 10 |
| Rod grounded | 56.5 | 1 | 80 |
|  |  | 10 | 30 |
|  |  | 100 | 10 |

It is apparent that while with a resistor value of 120 Ohms breakdown occurs 10 out of 10 times with a specific value of voltage, for a resistor value of 100 kOhms breakdown occurs only 1 out of 10 times with the same value of voltage. The results are influenced by the ground effect.

The results depend on the polarity of the applied voltage as well as the way of grounding (grounded plate or grounded rod).

It is also concluded from the experimental work (Fig. 8) that the in series connection of the resistor in the circuit has significant influence on the formation, the magnitude and the duration of the corona pulses (partial discharges) occurring in the gap before breakdown. The corona effects decrease as the resistor increases. Thus the peak value of the impulse voltage necessary for the breakdown (breakdown voltage) of the gap is increased. A new method of controlling the corona effects and consequently the breakdown voltage of small air gaps stressed by impulse voltage of short duration in connection to the ground effect and the polarity effect has arisen. The above method may be proven, after further investigation, to be applicable on many other insulating arrangements

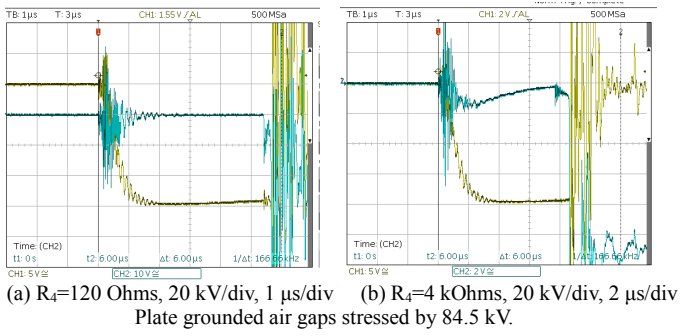

(a) $R_4$=120 Ohms, 20 kV/div, 1 μs/div    (b) $R_4$=4 kOhms, 20 kV/div, 2 μs/div
Plate grounded air gaps stressed by 84.5 kV.

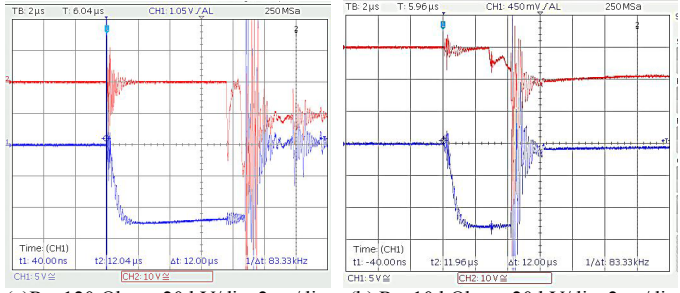

(a) $R_4$=120 Ohms, 20 kV/div, 2 μs/div    (b) $R_4$=10 kOhms, 20 kV/div, 2 μs/div
Rod grounded air gaps stressed by 56.5 kV.

Fig. 9 Oscillograms showing the corona pulses and the breakdown for the 10-100 mm rod-plate air gaps, with the plate grounded and a length of 4 cm, stressed by lighting impulse voltage.

### SIMULATION ANALYSIS RESULTS

P-Spice software has been used for simulation analysis. The equivalent model for the analysis is shown in Fig. 10.

As the applied voltage is a rapidly changing pulse of short duration, the air gaps capacitance presents a high capacitive reactance. Thus in the analysis model used, a need to add a resistor $R_5$ in series with the gap emerged, simulating with a high precision the reactance of the capacitance of the gap. The value of the capacitance of the gap ($C_{gap}$) was defined as well.

The addition of the above resistor $R_5$ in combination with the appropriate each time value of the capacitance, gave the following curves for voltage $V_{R4}$ as shown in Fig. 11, which match the relative curves of the oscillograms. Thus the value of the capacitance of the air gap is resulted from simulation analysis in connection to the experimental results.

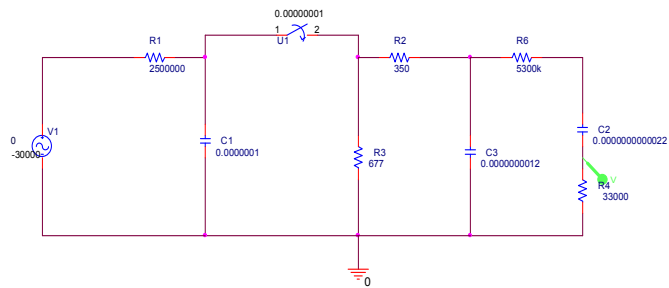

Fig. 10 Equivalent Model for the simulation analysis, with $R_4$= 33 kOhms, $R_5$=5,3 MOhms and $C_{gap}$=2,2 pF

It is resulted that for the circuit of Fig. 10, with $R_4$=33 kOhms and in order for the curves obtained from simulation to match the ones taken from experimental work the capacitance of the air gap arrangement should be set to 2.2 pF, and the resistor $R_5$ to 5.3 MOhms (Fig. 11).

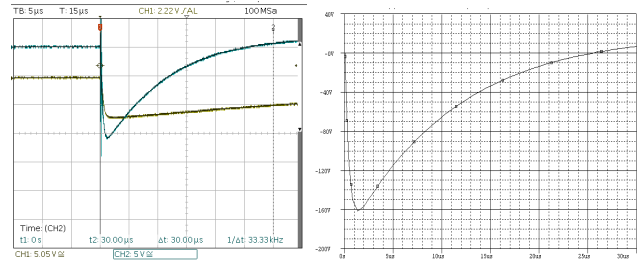

(a) 20 kV/div and 50 V/div, 2 μs/div    (b) 40 V/div, 5 μs/div

Fig. 11 Oscillograms from experimental arrangement with $R_4$=33 kOhms in series with the air gap. (a) From experimental arrangements 50 V/div, 5 μs/div, (b) from simulation analysis of the circuit of Fig. 9, with $R_5$=5.3 MOhms, and $C_{gap}$=2.2 pF.

### THEORETICAL - MATHEMATICAL ANALYSIS

Appropriate mathematical models of the experimental arrangements have been designed. The equation of the initial conditions concerning the analysis of the electric circuit and the calculation of the voltage of $C_2$, the electric current through $C_3$ and $R_4$, have been formulated. $C_3$ ($C_{gap}$) is the capacitance of the air gap.

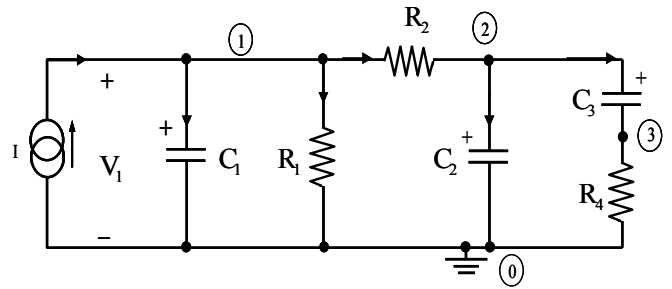

Fig. 12 Equivalent Model of the experimental set up for the mathematical analysis

The equivalent model for the theoretical analysis is the circuit of Fig. 12, and the equations for the initial conditions are:

From Kirckkoff's voltage law in loop $C_2$, $C_3$, $R_4$:

$$C_3 \frac{dV_{C_3}}{dt} R_4 + V_{C_3} = V_{C_2} \qquad (4)$$

From Kirckkoff's voltage law in loop $C_2$, $R_1$, $R_2$

$$i_{R_2} = \frac{V_{C_1} - V_{C_2}}{R_2} \qquad (5)$$

We apply Kirckkoff's current law in node 1:

$$C_1 \frac{dV_{C_1}}{dt} + \frac{V_{C_1}}{R_1} + \frac{V_{C_1} - V_{C_2}}{R_2} = I(t) \qquad (6)$$

From Kirckkoff's current law in node 2:

$$C_2 \frac{dV_{C_2}}{dt} = \frac{V_{C_1} - V_{C_2}}{R_2} - \frac{V_{C_2} - V_{C_3}}{R_4} \qquad (7)$$

Considering the state vector:

$$x = \begin{bmatrix} V_{C_1} & V_{C_2} & V_{C_3} \end{bmatrix}^T \qquad (8)$$

The state equations in matrix form are:

$$\frac{d}{dt}\begin{bmatrix}V_{C_1}\\V_{C_2}\\V_{C_3}\end{bmatrix}=\begin{bmatrix}-\frac{1}{C_1}\left(\frac{1}{R_1}+\frac{1}{R_2}\right) & \frac{1}{R_2 C_1} & 0\\ \frac{1}{R_2 C_2} & -\frac{1}{C_2}\left(\frac{1}{R_4}+\frac{1}{R_2}\right) & \frac{1}{R_4 C_2}\\ 0 & \frac{1}{R_4 C_3} & -\frac{1}{R_4 C_3}\end{bmatrix}\begin{bmatrix}V_{C_1}\\V_{C_2}\\V_{C_3}\end{bmatrix}+\begin{bmatrix}\frac{1}{C_1}\\0\\0\end{bmatrix}I(t) \quad (9)$$

The initial conditions are:

$$\begin{bmatrix}V_{C_1}(0)\\V_{C_2}(0)\\V_{C_3}(0)\end{bmatrix}=\begin{bmatrix}V_1\\0\\0\end{bmatrix} \quad (10)$$

The equations for the outputs are:

$$y=\begin{bmatrix}V_{C_3}\\i_{R_4}\end{bmatrix}=\begin{bmatrix}V_{C_3}\\ \frac{V_{C_2}-V_{C_3}}{R_4}\end{bmatrix}=\begin{bmatrix}0 & 0 & 1\\0 & \frac{1}{R_4} & -\frac{1}{R_4}\end{bmatrix}\begin{bmatrix}V_{C_1}\\V_{C_2}\\V_{C_3}\end{bmatrix} \quad (11)$$

Whereas for I(t)=0 and the given initial conditions:

$$(12)$$

$$\begin{bmatrix}V_{C_3}(t)\\i_{R_4}(t)\end{bmatrix}=\begin{bmatrix}0 & 0 & 1\\0 & \frac{1}{R_4} & -\frac{1}{R_4}\end{bmatrix}\exp\left\{\begin{bmatrix}-\frac{1}{C_1}\left(\frac{1}{R_1}+\frac{1}{R_2}\right) & \frac{1}{R_2 C_1} & 0\\ \frac{1}{R_2 C_2} & -\frac{1}{C_2}\left(\frac{1}{R_4}+\frac{1}{R_2}\right) & \frac{1}{R_4 C_2}\\ 0 & \frac{1}{R_4 C_3} & -\frac{1}{R_4 C_3}\end{bmatrix}t\right\}\begin{bmatrix}V_1\\0\\0\end{bmatrix}=$$

$$=\begin{bmatrix}0 & 0 & 1\\0 & \frac{1}{R_4} & -\frac{1}{R_4}\end{bmatrix}\exp\left\{\begin{bmatrix}-43342 & 28571.4 & 0\\ 2380952.4 & -2.380952.4-\frac{833333333.3}{R_4} & \frac{833333333.3}{R_4}\\ 0 & \frac{1}{R_4 C_3} & -\frac{1}{R_4 C_3}\end{bmatrix}t\right\}\begin{bmatrix}V_1\\0\\0\end{bmatrix}$$

Instead of the model of Fig. 12 a more convenient model as of Fig. 13 has been used, in which E is the voltage across the capacitor $C_2$.

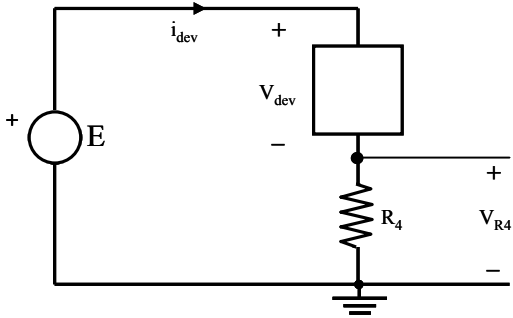

Fig. 13 Equivalent 2$^{nd}$ model used for the mathematical analysis

Values of charging current of the capacitance of the air gap arrangement were obtained by the oscilloscope's TRC data with the use of Matlab and were compared with the simulation results. The comparison results are impressive as shown in Fig. 14. The curves obtained from mathematical analysis match with the curves obtained from simulation.

With the same analysis the values of the capacitance of the air gap arrangement was calculated. The results are shown in Table 2.

The capacitance of the air gap can be calculated from equation (3). Using the data of the oscilloscope (TRC) the capacitance of the air gap arrangement $C_3=C_{gap}$ was calculated by the equation:

$$C_3 = T * \frac{\sum_{\lambda=1}^{k} i_{dev}(\lambda T)}{v_{dev}(kT)} \quad (13)$$

For the arrangement with $R_4$=33 kOhms the calculated value for the air gap capacitance results to 2.19 pF, and the resistor $R_5$ to 5.249 MOhms.

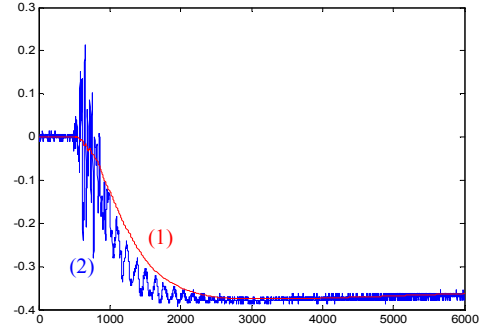

Fig. 13 Comparison of the simulation (1) and mathematical (2) graphs of the charging current of the capacitance of the air gap arrangement.

TABLE II
CALCULATION OF THE CAPACITANCE OF THE AIR GAP ARRANGEMENT, STRESSED BY 30 KV NEGATIVE IMPULSE VOLTAGE. $I_{CH}$ IS THE CHARGING CURRENT, $V_{CH}$ IS THE VOLTAGE ACROSS $R_4$, $C_3$ IS THE CAPACITANCE OF THE GAP.

| $R_4$ (kOhms) | 100 | 33 | 12 |
|---|---|---|---|
| Max(idev) (mA) | -1.7 | -4.8 | -11.5 |
| Time of gap charging, end $i_{ch}$ =0 (μs) | 47.3 | 25.8 | 11.3 |
| $C_3 = T * \frac{\sum_{\lambda=1}^{k} i_{dev}(\lambda T)}{v_{dev}(kT)}$ (pF) | 1.99 | 2.19 | 1.7 |
| $R_5$ (MOhms) | 16.04 | 5.25 | 1.81 |

The above calculated values are very close to the values from simulation analysis, and prove that the method is correct.

CONCLUSION

The V50% values of the lightning impulse breakdown voltage of small rod-plate air gaps are relatively higher than the breakdown voltage of the gaps stressed by dc voltage.

The ground effect in combination to the polarity effect influences the dielectric behavior of rod-plate air gaps stressed by impulse voltages in a different way than when stressed by dc voltage, due to the influence of the critical volume.

Higher values of the corona pulses and lower values of the breakdown voltages are recorded in the plate grounded arrangements, when stressed by positive voltage (rod positive). On the contrary lower values of the corona pulses and higher values of the breakdown voltages are recorded in the rod grounded arrangements when the plate is positively charged (rod negative). The differences, due to the polarity effect are bigger in the rod grounded than in the plate grounded arrangements, depending also on the rod's diameter.

When a resistor is connected in series to the air gap, the charging time ($t_{ch}$) of the air gap increases, while the peak voltage value ($V_{gap-peak}$) of the fully charged air gap decreases. Simultaneously and consequently the value of the breakdown voltage increases, while the corona pulses decrease.

A new method of controlling the dielectric behavior of an air gap stressed by impulse voltage arises.

Simulation analysis with the P-spice and mathematical analysis enable the calculation of the capacitance of the experimental air gap arrangement. The results of the simulation analysis are in full match with the results of mathematical analysis.

ACKNOWLEDGMENT

This research has been co-financed by the European Union (European Social Fund - ESF) and Greek national funds through the Operational Program "Education and Lifelong Learning" of the National Strategic Reference. Framework (NSRF) - Research Funding Program: ARCHIMEDES III. Investing in knowledge society through the European Social Fund.

Special thanks to Ass. Prof. John Parassidis, and Stefanos Zaoutsos, who offered considerable help in mathematical analysis.

REFERENCES

[1] Khalifa M., *High voltage engineering, theory and practice*, Marcel Dekker inc., New York, USA, 1990.
[2] Kuffel E., Zaengl W. S., Kuffel J., *High Voltage Engineering, Fundamentals,* Newnes Oxford, 2000.S.
[3] Naidu N. S., *High Voltage Engineering*, McGraw Hill, New York, 1996.
[4] Lazaridis L. A., Mikropoulos P. N., and Stasinopoulos C. A., "Breakdown in air and along a porcelain insulator under positive switching impulse voltages", ISH 15th Intern. Sympos. High Voltage Engineering, Ljubljana, Slovenia, T4-25061, pp. 154, 2007.
[5] Phillips D. B., Olsen R. G, and Pedrov P .D., "Corona Onset as a Design Optimization Criterion for High Voltage Hardware", IEEE Trans. Dielectr. Electr. Insul. Vol. 7 No 6, pp 744-751, December 2000.
[6] McAllister I. W., "Electric fields and electrical insulation", IEEE Trans. Dielectr. Electr. Insul, Vol. 9, pp. 672-696, 2002.
[7] Hua Wand Li J., Zhonggan Y., "The influence of ions on electrical breakdown in air", IEEE Proc. Dielectr. Materials, Vol. 1, pp. 168-171, 1988 S.
[8] F.V. Topalis, T. Katsabekis, P.D. Bourkas, I.A. Stathopulos: "Corona inception and breakdown in medium air gaps under switching impulse voltages", Proc. of IX Inter. Conf. on Gas Discharges and their Applications pp. 467-470, Venezia, September 1988.
[9] Michael M. H., S. Robert, "Partial discharge impulse characteristics of different detection systems", Technical report, Institute of High Voltage Engineering and System Management Graz University of Technology, Austria.
[10] Schwarz R., Muhr M., Pack S., "Evaluation of Partial Discharge Impulses with Optical and Conventional Detection Systems", Proc. of the XIVth International symposium on High Voltage Engineering, Tsinghua University , Beijing, China, August 25-29, 2005
[11] Maglaras A., Topalis F. V., "Influence of Ground and Corona Currents on Dielectric Behavior of Small Air Gaps", IEEE Trans. Dielectr. Electr. Insul, Vol. 16, No 1, pp. 32-41, 2009.
[12] Maglaras A., Topalis F. V., "The Influence of the Grounding of the Electrodes and the Corona Current to the Dielectric Behavior of the Air Gaps", Int. Conf. Med Power 08, Thessalonica, 2008.
[13] Maglaras L. A., Maglaras A.L., Frangiskos V. Topalis F. V., The influence of the Effect of Grounding and Corona Current on the Field Strength the Corona Onset and the Breakdown Voltage of Small Air Gaps, WSEAS TRANSACTIONS on POWER SYSTEMS, Issue 1, Volume 3, January 2008, ISSN: 1790-506.